\documentstyle[aps,prl,tighten,floats,epsfig]{revtex}

\begin{document}

\twocolumn[

\hsize\textwidth\columnwidth\hsize
\csname@twocolumnfalse\endcsname

\draft
\preprint{PLA about critical doping ($\tilde{x} \simeq  0.25$)}
\title{The overdoped regime in $La_{2-x}Sr_xCuO_4$}
\author{A.~Avella, F.~Mancini and D.~Villani$^\dag$}
\address{Universit\`a degli Studi di Salerno -- Unit\`a INFM di Salerno\\
Dipartimento di Scienze Fisiche ``E.R. Caianiello'', 84081 Baronissi,
Salerno, Italy}
\date{\today}
\maketitle

\begin{abstract}
Recent  experimental data for the overdoped $La_{2-x}Sr_xCuO_4$ have
firmly established the properties which characterize the transition
between the superconducting and the Fermi liquid metallic phases
($\tilde{x} \simeq 0.25$). The thermodynamic response functions show a
pronounced feature at this point, while the Fermi surface undergoes a
dramatic change. By use of the Composite Operator Method for the
two-dimensional Hubbard model, it is found that the presence of a van
Hove singularity in the lower Hubbard band can explain these behaviors.
\end{abstract}

\pacs{74.72.-h, 71.10.Fd, 71.10.-w}]

\narrowtext

In the last few years the availability of single-crystal samples for
overdoped $La_{2-x}Sr_xCuO_4$ ({\em LSCO}) has permitted an experimental
analysis \cite{Ino} of the critical point which signs the transition
between the superconducting and the Fermi liquid metallic phases
($\tilde{x} \simeq 0.25$) \cite{Phase,Susc}.

At this value of dopant concentration some thermodynamic properties
result to be greatly enhanced with a strongly pronounced peak. Indeed,
this is the case for the entropy \cite{Entro,Susc} and the uniform
static spin magnetic susceptibility \cite{Susc}. Further, at the same
doping, the Hall coefficient reverses its sign \cite{Hall} and an abrupt
change in the shape of the Fermi surface occurs \cite{Ino}. This kind of
behavior could be seen as a clear indication of the crossing of a van
Hove singularity through the chemical potential for that doping value
($\tilde{x} \simeq 0.25$) or, ultimately, of the relevance of band
dispersion driven processes in the overdoped region.

These experimental findings can be explained within an unique
theoretical framework where the electronic state evolution at large hole
doping is governed by a van Hove singularity. Beyond, this scenario is
only sensitive to the location of the peak and not to the overall band
structure.

\begin{figure}[tb]
\begin{center}
\epsfig{file=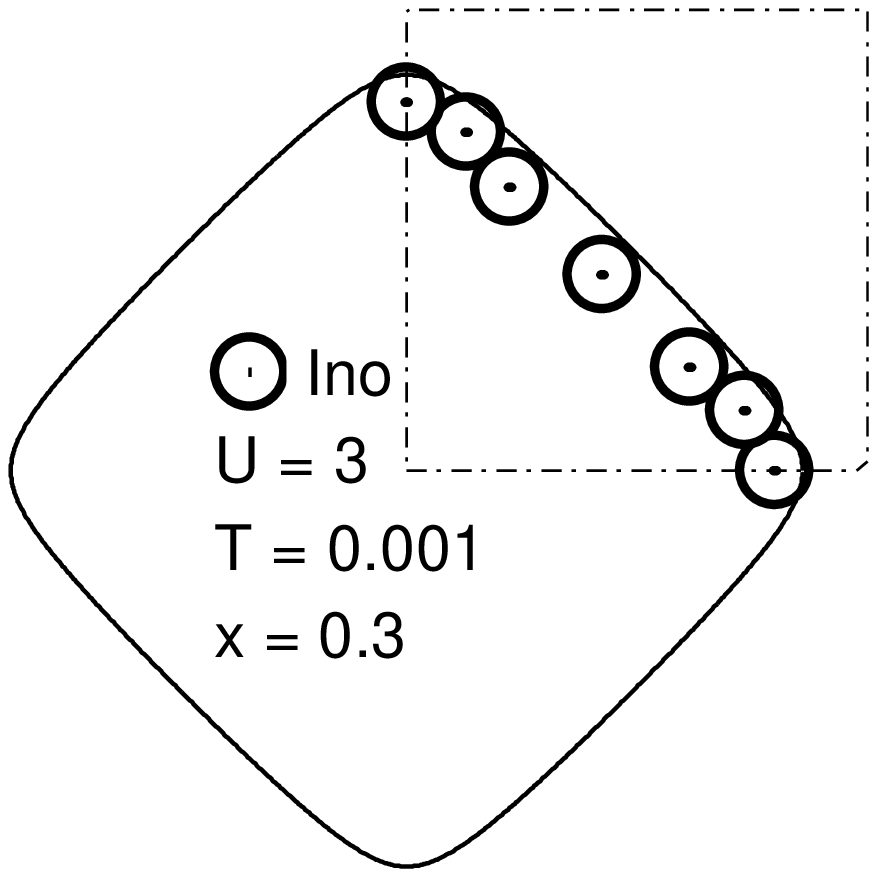,width=8cm,clip=}
\end{center}
\caption{Fermi surface for $U = 3$, $T = 0.001$ and $x = 0.3$.
The experimental data $\odot$ are taken from Ref.~\protect\cite{Ino}.}
\label{fig1}
\end{figure}

By means of the Composite Operator Method, in the two-pole approximation
\cite{Two}, we have studied the two-dimensional Hubbard model and, in
particular, the specific role played by the van Hove singularity in the
lower Hubbard band.

The 2D Hubbard model is defined as follows:
\begin{equation}
H=\sum_{ij}t_{ij}c^{\dag}(i)c(j)+U\sum_{i}n_\uparrow(i)n_\downarrow(i)
  -\mu\sum_{i}n(i)
\end{equation}
where $c(i)$ is the electron operator at the site $i$, in the spinor
notation. $t_{ij}$ denotes the transfer integral and describes hopping
between nearest neighbor sites; $U$ is the Hubbard interaction between
two $c$ electrons at the same site with $n_\sigma(i)$ being the number
operator per spin $\sigma$. $\mu$ is the chemical potential and $n(i)$
the total number operator.

As basic field for the causal thermal Green's function of the system we
have chosen the Hubbard doublet. This choice is dictated by the
equations of motion of the basic electronic field $c(i)$ and has the
advantage to automatically preserve the first four spectral momenta
\cite{SM}. In addition, it is known (quantum Monte Carlo and Lanczos
data) that due to the on-site Coulomb interaction two sharp features
develop in the band structure which correspond to the Hubbard subbands
and describe interatomic excitations mainly restricted to subsets of the
occupancy number \cite{Dag}. The use of a non-standard basis gives us
the possibility to implement a constraint on the solution with the
content of the Pauli principle \cite{Two}.

\begin{figure}[tb]
\begin{center}
\epsfig{file=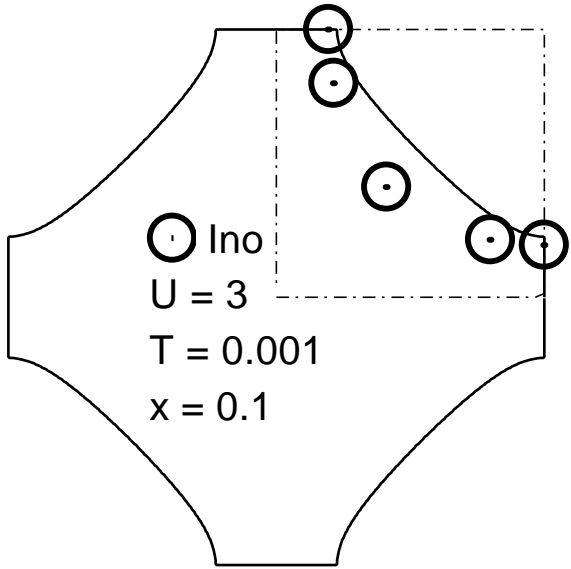,width=8cm,clip=}
\end{center}
\caption{Fermi surface for $U = 3$, $T = 0.001$ and $x = 0.1$.
The experimental data $\odot$ are taken from Ref.~\protect\cite{Ino}.}
\label{fig2}
\end{figure}

\begin{figure}[tb]
\begin{center}
\epsfig{file=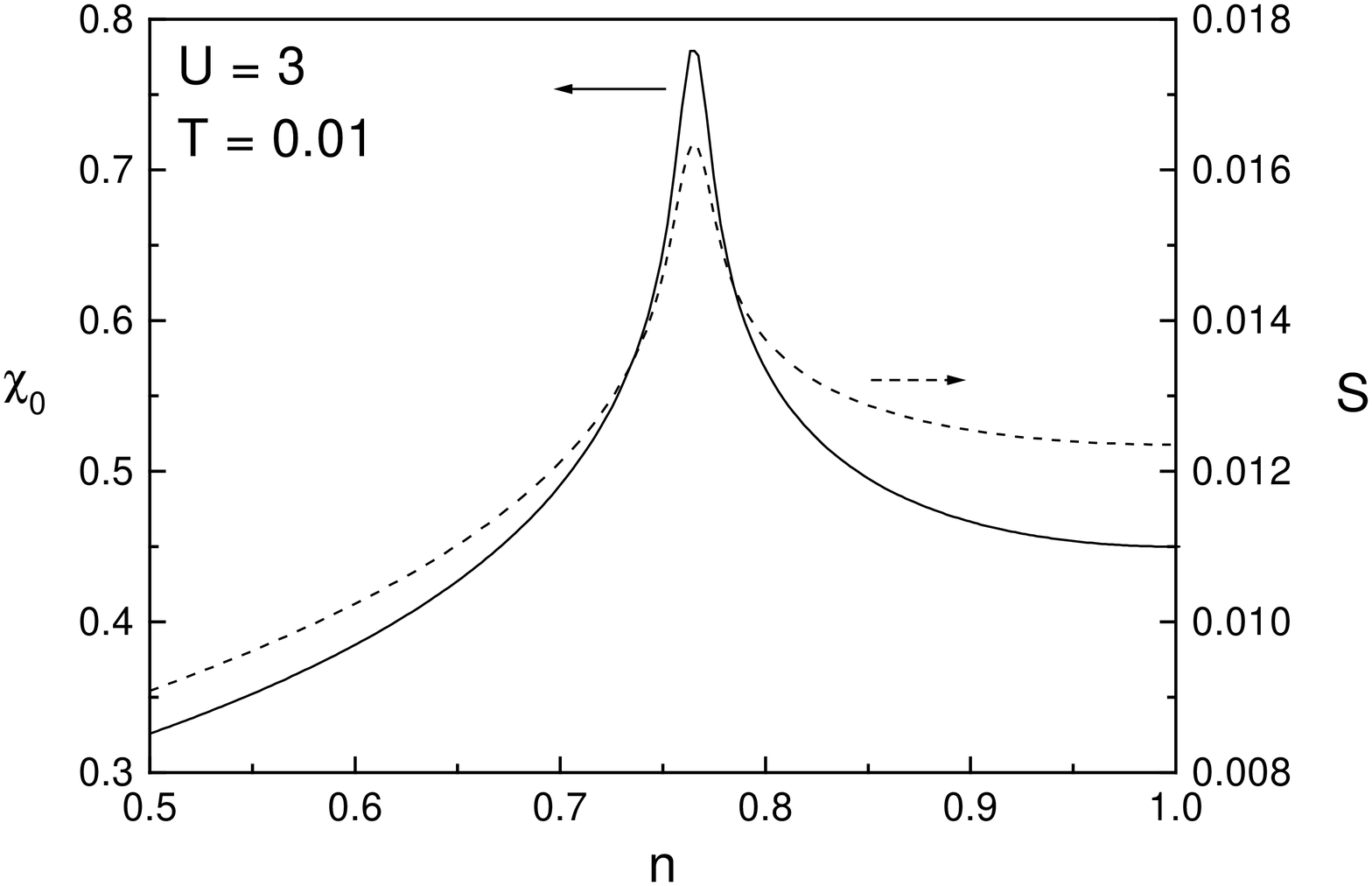,width=8cm,clip=}
\end{center}
\caption{The uniform static spin magnetic susceptibility $\chi_0$
and the entropy $S$ for $U = 3$, $T = 0.01$ as a function of the filling
$n$. The left and right vertical scale refer to the intensity of
$\chi_0$ and $S$, respectively.}
\label{fig3}
\end{figure}

The analytical expressions for the entropy, the uniform static spin
magnetic susceptibility and the Fermi surface have been already given in
Refs.~\cite{Thermo,Susce,Fermi}. The entropy has been calculated by
exploiting thermodynamic relations after the Maxwell's identities
\cite{Thermo}; the spin magnetic susceptibility has been determined by
use of the equations of motion \cite{Susce}. The Fermi surface is given
by the intersection between the lower Hubbard band and the energy level
corresponding to the chemical potential \cite{Fermi}.

The Fermi surface structure at $x = 0.3$ for the {\em LSCO} has been
only recently obtained by means of {\em ARPES} experiments \cite{Ino}.
The experimental data can be reproduced in the context of the
two-dimensional one-band Hubbard model with $U
= 3$, see Fig.~\ref{fig1}. Both the shape and the volume of the Fermi
surface qualitatively agree with experimental ones (errors of the order
of 15\%) showing that the nature of the metallic phase is of a
conventional Fermi liquid state. In fact, it should be considered as
experimentally robust that the destruction of the Fermi surface and the
relative issue of a novel state of the matter arise only in the
underdoped region \cite{FS}.

The comparison with the experimental results \cite{Ino} for $x = 0.1$,
see Fig.~\ref{fig2}, and $x_\textrm{c} = 0.15$ shows that the Fermi
surface changes its nature at a doping level between $0.15$ and $0.3$ in
agreement with the Hall coefficient experiments \cite{Hall}. The latter
set this critical doping at the same value at which the
superconductivity disappears \cite{Phase} and all the thermodynamic
properties related to the value of the density of states at the Fermi
level present a pronounced maximum ($\tilde{x} \simeq 0.25$)
\cite{Entro,Susc}. Our results for both the uniform static spin magnetic
susceptibility and the entropy show these features at the same doping
level ($\tilde{x} \simeq 0.25$) and for the same values of the model
parameters ($U = 3$), see Fig.~\ref{fig3}. A sharp maximum is well
developed and reproduces the experimentally observed one
\cite{Entro,Susc}. In particular, a pronounced maximum in the uniform
static spin magnetic susceptibility is a strong evidence for a
pseudo-nesting effect when accompanied by evidences of an ordinary Fermi
liquid phase.

\begin{figure}[tb]
\begin{center}
\epsfig{file=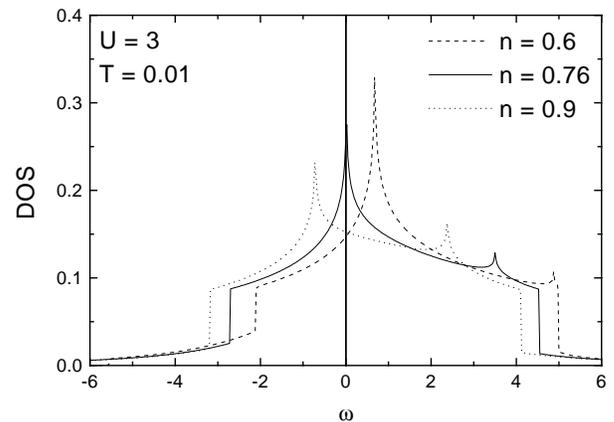,width=8cm,clip=}
\end{center}
\caption{Density of States for $U = 3$, $T = 0.01$ and different values of the filling $n$.}
\label{fig4}
\end{figure}

Now, analyzing the density of states for the same values of the model
parameters, see Fig.~\ref{fig4}, we can see that at the same doping
level ($\tilde{x} \simeq 0.25$) the lower Hubbard band van Hove
singularity crosses the chemical potential. It is worth to note that it
seems necessary to include second and third neighbor hopping to provide
both the dispersion and the line shape of the {\em ARPES} data for the
large variety of high-$T_\textrm{c}$ cuprates \cite{Kim}.

In conclusion, we have shown that the experimentally observed features
in the overdoped region can be described by the relative position of the
lower Hubbard band van Hove singularity. The latter crosses the Fermi
level at some critical doping ($\tilde{x} \simeq 0.25$) enhancing the
density of states and changing the Fermi surface nature.

\end{document}